\pdfoutput=1 
\documentclass[twocolumn,superscriptaddress,aps,preprintnumbers,amsmath,amssymb,prl,nofootinbib]{revtex4-2}

\usepackage{amsmath}
\usepackage{amssymb}
\usepackage{amsfonts}
\usepackage{graphicx}
\usepackage{xcolor}
\usepackage{xfrac}
\usepackage{comment}
\usepackage{pifont}
\usepackage{physics}
\usepackage{fourier}
\usepackage{hyperref}
\usepackage{bm}
\usepackage{enumitem}
\usepackage{tensor}

\definecolor{rossoferrari}{HTML}{D9073D}
\definecolor{mediumblue}{HTML}{0000CD}
\definecolor{forestgreen}{HTML}{228B22}
\definecolor{desy_blue}{HTML}{009EE2}
\definecolor{desy_orange}{HTML}{FD8800}
\definecolor{light_pink}{rgb}{1,0.4,0.4}
\definecolor{light_blue}{rgb}{0.284602,0.317763,0.963947}
\hypersetup{setpagesize=false,bookmarksnumbered=true,bookmarksopen=true,%
colorlinks=true,linkcolor=light_blue,urlcolor=rossoferrari,citecolor=rossoferrari,linktocpage=false}
\definecolor{dark_red}{rgb}{0.7, 0., 0.}
\definecolor{forestgreen}{HTML}{228B22}
\definecolor{ochre}{HTML}{CCAA2B}

\newcommand{\Mpl}{M_\mathrm{Pl}}
\newcommand{\bigzero}{\mbox{\normalfont\Large ${\bm 0}$}}
\newcommand{\rvline}{\hspace*{-\arraycolsep}\vline\hspace*{-\arraycolsep}}

\begin{document}


\preprint{KEK-TH-2547}
\preprint{KEK-Cosmo-0320}
\preprint{RESCEU-19/23}

\title{Geometry and unitarity of scalar fields coupled to gravity}

\author{Minxi He}
\email{heminxi@post.kek.jp}
\affiliation{Theory Center, IPNS, KEK, 1-1 Oho, Tsukuba, Ibaraki 305-0801, Japan}
\affiliation{Particle Theory and Cosmology Group, Center for Theoretical Physics of the Universe, Institute for Basic Science (IBS),  Daejeon, 34126, Korea}

\author{Kohei Kamada}
\email{kohei.kamada@resceu.s.u-tokyo.ac.jp}
\affiliation{Research Center for the Early Universe, The University of Tokyo, Hongo 7-3-1 Bunkyo-ku, Tokyo 113-0033, Japan}
\affiliation{School of Fundamental Physics and Mathematical Sciences, Hangzhou Institute for Advanced Study, University of Chinese Academy of Sciences (HIAS-UCAS),
Hangzhou 310024, China}
\affiliation{International Centre for Theoretical Physics Asia-Pacific (ICTP-AP), Beijing/Hangzhou, China}

\author{Kyohei Mukaida}
\email{kyohei.mukaida@kek.jp}
\affiliation{Theory Center, IPNS, KEK, 1-1 Oho, Tsukuba, Ibaraki 305-0801, Japan}
\affiliation{Graduate Institute for Advanced Studies (Sokendai), 1-1 Oho, Tsukuba, Ibaraki 305-0801, Japan}

\date{\today}


\begin{abstract}
\noindent
We formulate scalar field theories coupled non-conformally to gravity in a manifestly frame-independent fashion.
Physical quantities such as the $S$ matrix should be invariant under field redefinitions, and hence can be represented by the geometry of the target space.
This elegant geometric formulation, however, is obscured when considering the coupling to gravity because of the redundancy associated with the Weyl transformation.
The well-known example is the Higgs inflation, where the target space of the Higgs fields is flat in the Jordan frame but is curved in the Einstein frame.
Furthermore, one can even show that any geometry of O$(N)$ nonlinear $\sigma$ models can be flattened by an appropriate Weyl transformation.
In this Letter, we extend the notion of the target space by including the conformal mode of the metric, and show that the extended geometry provides a compact formulation that is manifestly Weyl-transformation/field-redefinition invariant.
We identify the cutoff scale with the inverse of square root of the extended target space curvature and confirm that it coincides with that obtained from two-to-two scattering amplitudes based on our formalism.

\end{abstract}


\maketitle


\noindent\textit{\textbf{Introduction.\,---\,}} %
Physical quantities should remain invariant under transformation between different descriptions of the same system.
Such redundancy has advantages in simplifying calculation.
However, it also unavoidably induces opacity of the consistency among different descriptions because the calculation can be drastically different.

Any scalar field theory in general is given the concept of geometry in its field-space manifold, or the target space, which is known as nonlinear $ \sigma $ model (NLSM)~\cite{Meetz:1969as,Honerkamp:1971sh,Honerkamp:1971xtx,Ecker:1971xko,Alvarez-Gaume:1981exa,Alvarez-Gaume:1981exv,Boulware:1981ns,Howe:1986vm}.
NLSMs frequently arise as low-energy effective field theories in various fields of theoretical physics.
They are particularly useful when the system undergoes some symmetry, whose information is encoded in the target space, \textit{e.g.,} the coset space for the effective theory of Nambu--Goldstone (NG) modes~\cite{Coleman:1969sm,Callan:1969sn}.
Since physical quantities such as the $S$ matrix should be invariant under field redefinitions, we may take whatever field basis of the target space so that the calculations become simple, at a cost of obscuring the field-redefinition invariance.
An elegant approach that makes this invariance manifest is formulating the physical observables geometrically since the geometry of target space is also invariant under field redefinitions~\cite{Dixon:1989fj,Alonso:2015fsp,Alonso:2016oah,Nagai:2019tgi,Cohen:2021ucp,Cheung:2021yog,Helset:2022tlf,Alonso:2023upf}.

However, such geometric formulation is spoiled once we turn on gravity.
The fundamental building block of gravitational theory is the metric of spacetime $g_{\mu\nu}$ through which all the fields couple to gravity.
Now we can perform the following redefinition of the metric $g'_{\mu\nu} = \Omega^2 g_{\mu\nu}$ with $\Omega$ an arbitrary function of matter fields, which is known as the Weyl or frame transformation.
It not only modifies the coupling to the Ricci scalar, but changes the geometry of the target space of an NLSM.
A famous example in cosmology is the Higgs inflation (HI)~\cite{Cervantes-Cota:1995ehs,Bezrukov:2007ep,Barvinsky:2008ia} which is originally defined in the Jordan frame where a large non-minimal coupling between the Standard Model (SM) Higgs fields and Ricci scalar is introduced to fit the observation of Cosmic Microwave Background~\cite{Planck:2018jri} while the target space of the Higgs fields is flat.
One may perform a Weyl transformation to the Einstein frame~\cite{Maeda:1988ab} where the Higgs fields are minimally coupled but the target space is curved.
This redundancy makes the determination of the target-space geometry ambiguous.

As an NLSM is merely a low-energy effective field theory (EFT), its validity is restricted up to a certain energy scale, corresponding to the target-space curvature.
This scale $ \Lambda $ can be extracted as the unitarity-violation scale of tree-level scattering amplitudes, and again the geometric formulation provides a powerful toolkit~\cite{Alonso:2015fsp,Nagai:2019tgi,Cohen:2021ucp,Alonso:2023upf}.
Along the same way, to verify the validity of HI as an EFT for inflation, the inflation scale needs to stay below $ \Lambda $.\footnote{
	The momenta of the gauge bosons produced during preheating can exceed the low cutoff scale so the system becomes strongly coupled~\cite{Ema:2016dny}.
}
However, owing to the presence of gravity, the geometric formulation is not applicable, and we had
to calculate all relevant processes in different frames and compare the results to guarantee consistency~\cite{Hertzberg:2010dc,Bezrukov:2010jz,Bezrukov:2011sz,Ren:2014sya,Ema:2019fdd,Hill:2020oaj,Ito:2021ssc,Karananas:2022byw}, which has caught intensive attention and debate due to the ambiguity (see also~\cite{Burgess:2009ea,Barbon:2009ya}).

In this Letter, we extend the notion of target-space geometry to preserve the advantages of the geometric approach even with gravity.
Specifically, we include the conformal mode of spacetime metric $\det (g_{\mu\nu})$ as a coordinate of the target space~\cite{Ema:2020zvg,Ema:2020evi}.
The extended target-space geometry is manifestly invariant even under the Weyl transformation because it corresponds to redefining the conformal mode.
We provide the geometrical meaning of the cutoff scale, which is manifestly independent of the frame and state.

\smallskip\noindent\textit{\textbf{Metric Higgs inflation.\,---\,}} %
We begin our discussion by considering HI in a metric formulation of gravity, which is based on the following action:
\begin{equation}\label{eq-metric-jordan}
	S = \int \dd^4 x \sqrt{- g_\text{J}}\,
	\qty( \frac{\Mpl^2 + \xi \phi^2}{2} R_\text{J} - \frac{g^{\mu\nu}_\text{J}}{2} \delta_{ij} \partial_\mu \phi^i \partial_\nu \phi^j - V ).
\end{equation}
Here $g_{\text{J}\mu\nu}$ is the metric of spacetime in the Jordan frame with its determinant being $g_\text{J}$,
$R_\text{J}$ is the Ricci curvature determined by $g_{\text{J} \mu\nu}$,
$\Mpl$ is the reduced Planck mass,
$\phi^i$ is a multi-component scalar field whose index $i$ runs through $i = 1, \cdots, N$ with $N \geqslant 2$, $\phi^2\equiv\delta_{ij}\phi^i \phi^j$,
and $V$ is a potential of the scalar field invariant under O$(N)$ rotation.
The target space spanned by $ \phi^i $ is trivial, \textit{i.e.} $ \delta_{ij} $.
By identifying $\phi^i$ as the SM Higgs doublet for $N = 4$ and $V$ as the SM Higgs potential~\cite{Cervantes-Cota:1995ehs,Bezrukov:2007ep,Barvinsky:2008ia}, one can show that the large expectation value of the SM Higgs fields exhibits the cosmic inflation perfectly consistent with observations~\cite{Planck:2018jri}.
In this case, the SM Higgs fields couple to \textit{e.g.}, gauge bosons which acquire mass terms for a finite vacuum expectation value (VEV) of Higgs fields. 
In the discussion of unitarity, we are interested in the behavior at a higher energy than the VEV.
For this reason, only the longitudinal modes are important and thereby the Goldstone equivalence theorem guarantees that our action \eqref{eq-metric-jordan} is sufficient~\cite{Burgess:2010zq,Bezrukov:2011sz,Bezrukov:2010jz,Gorbunov:2018llf,Karananas:2022byw}.

We can write down the same model in a seemingly different form by performing the following Weyl transformation
\begin{equation}\label{eq-weyl-tr-metric}
	g_{\text{E}\mu \nu} = \Omega^2 g_{\text{J}\mu\nu}, \qquad \Omega^2 \equiv 1 + \frac{\xi \phi^2}{\Mpl^2},
\end{equation}
which leads to the action in the Einstein frame 
\begin{equation}\label{eq-metric-einstein}
	S = \int \dd^4 x \sqrt{- g_\text{E}}\,
	\qty(  \frac{\Mpl^2}{2} R_\text{E}
	- \frac{g^{\mu\nu}_\text{E}}{2} G_{ij}^\text{E} \partial_\mu \phi^i \partial_\nu \phi^j - \frac{V}{\Omega^4} ),
\end{equation}
where the target-space metric is given by
\begin{equation}
	G_{ij}^\text{E} \equiv \frac{1}{\Omega^2} \qty( \delta_{ij} + \frac{6 \xi^2}{\Omega^2} \frac{\phi_i \phi_j}{\Mpl^2} ).
\end{equation}
One may readily confirm that the target space of $\phi^i$ is curved in this frame although it is flat in the Jordan frame \eqref{eq-metric-jordan}.
Nevertheless, the physical quantities should be unchanged as the Weyl transformation \eqref{eq-weyl-tr-metric} is merely a field redefinition.

This observation motivates us to extend the notion of the target space so that the extended geometry is also invariant under the Weyl transformation.
For this purpose, we extract the conformal mode of the metric as~\cite{Ema:2020zvg,Ema:2020evi}\footnote{Here, we focus on the discussion of frame-independence so only the conformal mode is sufficient. One might need to include the full spacetime metric as in Ref.~\cite{Finn:2019aip} if more comlicated transformation is involved such as the disformal transformation that mixes the graviton components and scalar fields.}
\begin{equation}\label{eq-conformal-mode}
	g_{\bullet \mu\nu} = \frac{\Phi_\bullet^2}{6 \Mpl^2} \tilde g_{\mu\nu} , \qquad \det \qty(\tilde g_{\mu\nu}) = -1,
\end{equation}
where the black dot implies a subscript associated with the frame, \textit{e.g,} $\bullet = \text{J}, \text{E}$.
The Weyl transformation of Eq.~\eqref{eq-weyl-tr-metric} now turns into the field redefinition of
\begin{equation} \label{eq-weyl-conformal-mode}
	\Phi_\text{E}^2 = \Omega^2 \Phi_\text{J}^2.
\end{equation}
Once the target space is extended to involve $\Phi_\bullet$, its geometry is manifestly invariant under not only the field redefinition of $\phi_i$ but the Weyl transformation.

We rewrite the action~\eqref{eq-metric-jordan} in a field basis of $(\varphi^a_\text{J}) = (\Phi_\text{J}, \phi^i)$.
A straightforward calculation leads to the following action:
\begin{equation}\label{eq-metric-jordan-conformal}
	S = \int \dd^4 x\, \qty(
		\frac{\Phi_\text{J}^2}{12} \Omega^2 \tilde R
		- \frac{ \tilde g^{\mu\nu} }{2} G_{ab}^{\text{J}}\partial_\mu \varphi_\text{J}^a \partial_\nu \varphi_\text{J}^b
		- \frac{\Phi_\text{J}^4 V}{36 \Mpl^4}
	),
\end{equation}
where the extended target-space metric reads
\begin{equation}\label{eq-field-space-jordan}
	\qty(G_{ab}^\text{J} )
	\equiv \begin{pmatrix}
		-\Omega^2 & -\xi \Phi_\text{J} \phi_j/\Mpl^2 \\
		-\xi \Phi_\text{J} \phi_i/\Mpl^2 & \frac{\Phi_\text{J}^2}{6 \Mpl^2} \delta_{ij}
	\end{pmatrix}.
\end{equation}
Here the Ricci curvature $\tilde R$ is given by $\tilde g_{\mu\nu}$. Physical quantities should be represented by the geometry specified by $G_{ab}^\text{J}$.

\begin{table*}[t]
	\caption{Target-space Riemann tensor at $(\bar \varphi^a_\text{J}) = (\sqrt{6}\Lambda_\text{G}/\bar{\Omega}, v, 0, \cdots, 0)$ in several examples.
	The field vector is expanded as $(\varphi^a_\text{J}) = (\bar \varphi^a_\text{J}) + (\delta\Phi_\text{J}, h, \pi^1, \cdots, \pi^{N-1})$.
	Note that the results are independent of frames so we drop the frame index.
	$\bar R_\text{others}$ means that at least one of the indices is $\delta \Phi_\text{J}$.
	We parametrize the Riemann tensor as
	$\bar R_{ikjl} = \bar R_{ik}( \delta_{ij} \delta_{kl} - \delta_{il} \delta_{jk} )$.
	}
	\begin{tabular}{l||c|c|c}
		& Metric HI & Einstein--Cartan HI & NLSM \\
		\hline
		$ \bar{R}_{h\pi^k} $
		& $\frac{(1 + 6 \xi )^2 \Lambda_\text{G}^2 }{6 (\Mpl^2 + \xi v^2 ) [\Mpl^2 + (1 + 6 \xi) \xi v^2 ]} $
		& $ \Lambda_\text{G}^2 \frac{[ \xi (1+6r^2\xi)v^2 +(1+6\xi)\Mpl^2 ]^2-36(1-r^2)\xi^2\Mpl^4}{6(\Mpl^2+\xi v^2)^3 [ \Mpl^2 +\xi (1+6r^2\xi)v^2 ]} $
		&  $ \Lambda_G^2
		\frac{
				\bar{G}^2 + 12 \Mpl^2 \bar{G} (\overline{f'} + v^2 \overline{f''}) + 6\Mpl^2
				\{
					\bar f \overline{G'}
					+ \overline{f'}
					[6 \Mpl^2
						(\overline{f'} + 2 v^2 \overline{f''})
						- v^2 \overline{G'}
					]
				\}
			}
			{
				6\Mpl^4 \bar{f}^2(\bar{G}+6\Mpl^2 v^2 \overline{f'}^2 \bar{f}^{-1})
			} $
        \\[0.8em]
		$ \bar{R}_{\pi^i \pi^k} $
		& $\frac{(1 + 6 \xi)^2 \Lambda_\text{G}^2 }{6 (\Mpl^2 + \xi v^2) [\Mpl^2 + (1 + 6 \xi) \xi v^2 ]} $
		& $ \Lambda_\text{G}^2 \frac{\xi(1+6\xi)(1+6r^2\xi)v^2 +[1+12\xi (1+3r^2\xi)]\Mpl^2}{6(\Mpl^2+\xi v^2)^2 [ \Mpl^2 +\xi (1+6r^2\xi)v^2 ] } $
		& $ \Lambda_G^2 \frac{ (6\Mpl^2\bar{f} + v^2) (\bar{G}-1) +v^2 ( 1+6\Mpl^2 \overline{f'})^2 }{6\Mpl^4 v^2 \bar{f}^2(\bar{G}+6\Mpl^2 v^2 \overline{f'}^2 \bar{f}^{-1})} $ \\[0.8em]
		$ \bar{R}_\text{others}$
		& $0$
		& $0$
		& $0$
	\end{tabular}
	\label{tab-riemann-tensor}
\end{table*}

Let us estimate the cutoff scale of this theory in the geometric language with the extended target space.
Hereafter we assume the contribution from the potential is subdominant so we can drop it. This is actually true for HI whose potential is up to quartic order with sufficiently small coupling.
One convenient way of extracting the cutoff scale is to consider the two-to-two scattering of $\phi^i$ as it grows in proportional to $E^2_\text{CM}/\Lambda_\text{G}^2 $ with $E_\text{CM}$ being the center-of-mass energy.
$ \Lambda_\text{G} $ is the unavoidable UV cutoff scale above which the graviton becomes strongly coupled, and to which the ratio of dimensionful quantities acquires physical meaning, usually chosen as $ \Mpl $.
Thus, one can extract the cutoff scale by requiring each amplitude to be smaller than unity.
Consider scatterings around the background of $(\bar \varphi^a_\text{J}) = (\sqrt{6}\Lambda_\text{G}/\bar{\Omega}, v, 0, \cdots, 0)$\footnote{
		We choose $ \bar{\Omega}(v) \bar{\Phi}_\text{J} = \sqrt{6} \Lambda_\text{G} $ as background such that the kinetic term of graviton is expressed as $\Lambda_\text{G}^2 (\partial_\rho h_{\mu\nu})^2 / 8$ with $\tilde g_{\mu\nu} = \eta_{\mu\nu} + h_{\mu\nu}$.
		}
and $\bar{\tilde{g}}_{\mu\nu} = \eta_{\mu\nu}$, which is relevant for the unitarity violation during HI.
Geometric language provides the following elegant expression of general four-point amplitudes~\cite{Nagai:2019tgi,Cohen:2021ucp,Cheung:2021yog,Alonso:2023upf}
\begin{align} \label{eq-four-point}
	&\mathcal{M}_{I J \leftrightarrow K L} = \frac{2}{3} \qty[
		s_{IJ} \bar R_{I (K L) J} + s_{IK} \bar R_{I (J L) K}  + s_{IL} \bar R_{I (J K) L}
	],
\end{align}
with $s_{IJ} \equiv (p_I + p_J)^2$.
The subscripts $I$ specify the states, where the capital letter indicating that the states should be canonically normalized, \textit{e.g.,} $I = H$ for the Higgs mode.
The relation between a field basis $\varphi^a$ and the canonically normalized states at $\bar\varphi^a$ are provided by the vielbein, \textit{e.g.}, $ \bar G_{ab}^{\rm J} = \bar e_a^{~A} \bar e_b^{~B} \eta_{AB} $.
The parentheses in the subscripts denote the symmetrization.
The Riemann tensor is
\begin{align}\label{eq-two-Riemann}
	\bar R_{ABCD} = \bar e_A^{~a} \bar e_B^{~b} \bar e_C^{~c} \bar e_D^{~d} \bar R_{abcd} ,
\end{align}
with the vielbein and Riemann tensor being evaluated at $\bar\varphi^a$.
To estimate the scattering amplitudes among Higgs $H$ and NG bosons $\Pi^i$ at $\bar\varphi^a_\text{J}$, all we need is the following vielbein:
	\begin{equation} \label{eq-shift-diagonalization}
		\qty(\bar e_{A}^{~a}) = \frac{\Mpl \bar\Omega}{\Lambda_\text{G}}
		\begin{pmatrix}
			\begin{matrix}
				\frac{\Lambda_\text{G}}{\Mpl \bar\Omega^2} & -\frac{\sqrt{6} \xi \Lambda_G v / \Mpl^2}{\bar\Omega^3 \sqrt{1+ 6\xi^2 v^2/(\Mpl^2 \bar\Omega^2)}} \\
				0 & \frac{1}{\sqrt{1+ 6\xi^2 v^2/(\Mpl^2 \bar\Omega^2)}}
			\end{matrix}
			&\rvline &  \bigzero \\
			\hline
			\bigzero & \rvline & \mathbb{1}
		\end{pmatrix}
		.
	\end{equation}

In the literatures, \textit{e.g.,} \cite{Ren:2014sya,Ema:2019fdd,Karananas:2022byw,Ito:2021ssc}, the scattering amplitudes among the Higgs and NG bosons are computed explicitly to confirm the invariance between the Einstein and Jordan frames.
Owing to the extended geometry of the target space, the invariance under the frame transformation with fixed incoming and outgoing states now becomes manifest.
This is because the Weyl transformation is a particular coordinate transformation with respect to the field indices $a$ in the extended target space, which are already contracted as given in Eq.~\eqref{eq-two-Riemann}.

Furthermore, the physical quantities such as the cutoff scale should
not even depend on the choice of incoming and outgoing states.
This motivates us to consider the Ricci scalar of the extended target space, which is frame, coordinate, and states independent,
\begin{align}\label{eq-ricci-scalar-special}
	\bar R = \bar G^{ab} \bar G^{cd} \bar R_{acbd}= 2(N-1) \qty( \bar R_{H \Pi^k} + \frac{N-2}{2} \bar R_{\Pi^i \Pi^k} ) .
\end{align}
In the second equality, we have used $\bar R_{ABCD} \neq 0$ only if all the indices are $I = H, \Pi^k$, and $\bar R_{IKJL} = \bar R_{IK} (\delta_{IJ} \delta_{KL} - \delta_{IL} \delta_{JK})$, which follows from Eq.~\eqref{eq-shift-diagonalization} and Tab.~\ref{tab-riemann-tensor}.
The cutoff scale of the theory is then given by $ \Lambda_\text{metric}/ \Lambda_\text{G} \sim \sqrt{N^2/\bar R} /\Lambda_\text{G} $.

Now we confirm that the cutoff scale extracted from the scattering amplitudes coincides with the Ricci scalar by explicit computations.
From Tab.~\ref{tab-riemann-tensor} and Eq.~\eqref{eq-shift-diagonalization}, the nonvanishing scattering amplitudes read
\begin{align} \label{eq-pipipipi-metric}
	\mathcal{M}_{\Pi^i \Pi^i \leftrightarrow \Pi^j \Pi^j}
	&=
	- \frac{s_{12}}{6 \Lambda_\text{G}^2} \frac{\qty(1 + 6 \xi)^2 ( \Mpl^2 + \xi v^2 )}{\Mpl^2 + (1 + 6 \xi) \xi v^2 } \quad \text{for}~~ i \not = j, \\
	\label{eq-highpipi-metric}
	\mathcal{M}_{HH \leftrightarrow \Pi^i \Pi^i}
	&=
	- \frac{s_{12}}{6 \Lambda_\text{G}^2} \frac{(1 + 6 \xi)^2 ( \Mpl^2 + \xi v^2 )^2}{\qty[\Mpl^2 + (1 + 6 \xi) \xi v^2]^2}.
\end{align}
Consequently, we obtain the cutoff scale of the metric HI
\begin{equation}\label{eq-metric-cutoff-pipi}
	\frac{\Lambda_\text{metric}}{\Lambda_\text{G}} \sim \frac{\bar R_{\Pi^i \Pi^k}^{-1/2}}{\Lambda_\text{G}} \sim
	\begin{cases}
		1/\xi & \text{for} \quad v \lesssim \Mpl / \xi, \\
		v/\Mpl & \text{for} \quad \Mpl / \xi \lesssim v \lesssim \Mpl/\sqrt{\xi}, \\
		1/\sqrt{\xi} &\text{for} \quad \Mpl/\sqrt{\xi} \lesssim v,
	\end{cases}
\end{equation}
for $N > 2$, and
\begin{equation}\label{eq-metric-cutoff-hpi}
	\frac{\Lambda_\text{metric}}{\Lambda_\text{G}} \sim \frac{\bar R_{H \Pi^k}^{-1/2}}{ \Lambda_\text{G}} \sim
	\begin{cases}
		1/\xi & \text{for} \quad v \lesssim \Mpl / \xi, \\
		\xi v^2 / \Mpl^2 & \text{for} \quad \Mpl / \xi \lesssim v \lesssim \Mpl / \sqrt{\xi}, \\
		1 & \text{for} \quad \Mpl / \sqrt{\xi} \lesssim v,
	\end{cases}
\end{equation}
for $N = 2$ where we do not have $\Pi^i \Pi^i \leftrightarrow \Pi^j \Pi^j$ scattering.
Our results are consistent with the literature such as Refs.~\cite{Bezrukov:2010jz,Bezrukov:2011sz,Ren:2014sya,Mikura:2021clt,Ito:2021ssc,Karananas:2022byw}.
It is clear that the obtained cutoff scale indeed coincides with the Ricci scalar given in Eq.~\eqref{eq-ricci-scalar-special}, which is frame and state independent.
We emphasize that the physically relevant quantity is the ratio $\Lambda_\mathrm{metric}/\Lambda_\mathrm{G}$, which is frame independent.

\smallskip\noindent\textit{\textbf{Other Higgs inflation.\,---\,}} %
Our discussion is also applicable to alternative formalisms of gravity like Palatini HI~\cite{Bauer:2010jg,Rasanen:2018ihz}. It is recently realized that HI in the Einstein--Cartan gravity with a non-minimally coupled Nieh--Yan term~\cite{Nieh:1981ww} can serve as a general setup that includes the well-known metric and Palatini cases~\cite{Shaposhnikov:2020gts}, so here we consider Einstein--Cartan HI for a general discussion. In Einstein--Cartan HI,
the affine connection $ \Gamma_{\mu\nu}^{\rho} $ is treated \textit{a priori} independently of
$ g_{\mu\nu} $,
although the action is of the same form as Eq.~\eqref{eq-metric-jordan}. The non-minimally coupled Nieh--Yan term is~\cite{Shaposhnikov:2020gts}
\begin{equation}
    -\frac{\xi_\eta}{4} \int \dd^4 x \, \phi_i^2 \partial_\mu \left( \epsilon^{\mu\nu\rho\sigma} T_{\nu\rho\sigma} \right),
\end{equation}
where $ \xi_{\eta} $ is the coupling constant, $ \epsilon^{\mu\nu\rho\sigma} $ is the Levi-Civita symbol such that $ \epsilon^{0123}=1 $,
and $ T_{~\mu\nu}^{\rho} \equiv \Gamma_{\mu\nu}^{\rho} - \Gamma_{\nu\mu}^{\rho} $ is the torsion tensor.
One can solve the constraint equation for $ \Gamma $ to obtain an equivalent action in the same form as Eq.~\eqref{eq-metric-jordan-conformal}
but $ G_{ab}^\text{J} $ is replaced by~\cite{He:2023vlj}
\begin{equation}
	\qty(G_{ab}^\text{EC} )
	\equiv \begin{pmatrix}
		-\Omega^2 & -\xi \Phi_\text{J} \phi_j/\Mpl^2 \\
		-\xi \Phi_\text{J} \phi_i/\Mpl^2 & \frac{\Phi_\text{J}^2}{6 \Mpl^2} \qty(\delta_{ij}-\frac{6(1-r^2)\xi^2}{\Mpl^2\Omega^2}\phi_i \phi_j)
	\end{pmatrix},
\end{equation}
where we have defined $ r\equiv \xi_{\eta}/\xi $.
Thus,
$ r=1 $ recovers the results in the previous section while $ r =0 $ reproduces the Palatini HI.

Following previous procedures, the frame-independent cutoff scale is obtained by calculating the Ricci scalar for the extended target space, whose structure is the same as Eq.~\eqref{eq-ricci-scalar-special}:
$\Lambda_\text{EC}/\Lambda_\text{G} \sim \bar R^{-1/2} / \Lambda_\text{G}$.
For
$ 1/\sqrt{\xi} \lesssim r \leqslant 1 $,
\begin{equation}
	\frac{\Lambda_\text{EC}}{\Lambda_\text{G}} \sim
	\begin{cases}
		1/(r\xi) & \text{for} \quad v \lesssim \Mpl / (r\xi), \\
		v/\Mpl & \text{for} \quad \Mpl / (r \xi) \lesssim v \lesssim \Mpl/\sqrt{\xi}, \\
		1/\sqrt{\xi} &\text{for} \quad \Mpl/\sqrt{\xi} \lesssim v,
	\end{cases}
\end{equation}
for $N > 2$, and
\begin{equation}
	\frac{\Lambda_\text{EC}}{\Lambda_\text{G}} \sim
	\begin{cases}
		1/(r\xi) & \text{for} \quad v \lesssim \Mpl / (r \xi), \\
		r \xi v^2/\Mpl^2 & \text{for} \quad \Mpl / (r \xi) \lesssim v \lesssim \Mpl / \sqrt{\xi}, \\
		r\sqrt{\xi} v/\Mpl & \text{for} \quad \Mpl / \sqrt{\xi} \lesssim v \lesssim \Mpl / (r\sqrt{\xi}), \\
		1 & \text{for} \quad \Mpl / (r\sqrt{\xi}) \lesssim v,
	\end{cases}
\end{equation}
for $N = 2$.
As for $ 0\leqslant r \lesssim 1/\sqrt{\xi} $, the cutoff is basically $ \sim 1/\sqrt{\xi} $ except for $ v \gtrsim \Mpl/\sqrt{\xi} $ in the $ N=2 $ case where $ \Lambda_\text{EC}/\Lambda_\text{G} \sim v/\sqrt{v^2+12\Mpl^2} $. These results are all frame independent.

\smallskip\noindent\textit{\textbf{General nonlinear $\sigma$ model.\,---\,}} %
We can easily apply our approach to general NLSM with gravity. Consider \textit{e.g.,} multiple scalars with curved target space and non-minimal coupling in metric formalism\footnote{We do not consider an arbitrary function $ F(\phi^i, R) $ in the gravity part simply to avoid the complication from a new scalar degree of freedom which generically appears in $ f (R) $ theories in metric formalism and is beyond the purpose of this letter.}
\begin{equation}
	S =\!\! \int \! \! \dd^4 x \sqrt{- g_\text{J}}\,
	\qty( \frac{\Mpl^2}{2} f R_\text{J} - \frac{g^{\mu\nu}_\text{J}}{2} \mathcal{G}_{ij} \partial_\mu \phi^i \partial_\nu \phi^j - V ),
\end{equation}
where $ f $ is a positive-definite scalar function with $f = 1$ for $\phi^i = 0$ and $ \mathcal{G}_{ij} $ is a general non-degenerate target-space metric both of which depend on $ \phi^i $.
We further assume the theory respects O$ (N) $ symmetry ($ N \geqslant 2 $) as a simple example for NLSM.
This symmetry allows us to rewrite the metric as
$\mathcal{G}_{ij} \dd \phi^i \dd \phi^j = G (h^2) \dd h^2 + (h^2/v^2) [ \dd \vec{\pi}^2 + (\vec{\pi}\cdot \dd \vec{\pi})^2 / (v^2 - \vec{\pi}^2) ]$ with $G(0) = 1$ in the spherical coordinate of $ \phi^i = (h, \vec{\pi})$,
and restricts the form of the non-minimal coupling to be $ f=f(h^2) $.
Here, $ h $ is the radial mode and $ \vec{\pi} $ are the coordinates on $ S^{N-1} $. $v$ is a parameter to give a mass dimension one to $\vec{\pi}$.
For $h = v$ and $\vec{\pi} = \vec{0}$, the Riemann tensor of $\mathcal{G}_{ij}$, denoted as $\mathcal{R}_{ikjl}$, is readily obtained as
$\bar{\mathcal{R}}_{\pi^i \pi^k} = (\bar G - 1) / (v^2 \bar G)$ and
$\bar{\mathcal{R}}_{h \pi^k} = \overline{G'} / \bar G$ with $\bar G = G(v^2)$ and $\overline{G'} = G'(v^2)$.
We again parametrize the Riemann tensor as $\bar{\mathcal{R}}_{ikjl} = \bar{\mathcal{R}}_{ik} (\delta_{ij}\delta_{kl} - \delta_{il}\delta_{jk})$.
One may confirm the restoration of O$(N)$ symmetry in the limit of $v \to 0$ as $\bar{\mathcal{R}}_{\pi^i \pi^k} = \bar{\mathcal{R}}_{h \pi^k} = \overline{G'_0}$ with $\overline{ G'_0 } = G'(0)$.

The extended target-space metric for $ \varphi^a_\text{J} = (\Phi_\text{J}, h, \vec{\pi}) $ is given as
\begin{equation}
    \qty(G_{ab}^\text{NL} )
	\equiv \begin{pmatrix}
		\begin{matrix}
			-f & - f' \Phi_\text{J} h \\
		- f' \Phi_\text{J} h  & \frac{\Phi_\text{J}^2}{6 \Mpl^2} \mathcal{G}_{hh}
		\end{matrix}
		&\rvline & \bigzero \\[1.5em]
		\hline
		\bigzero & \rvline & \frac{\Phi_\text{J}^2}{6 \Mpl^2} \qty( \mathcal{G}_{\pi^i \pi^j} )
	\end{pmatrix}.
\end{equation}
The relevant components of the target-space metric and Riemann tensor are shown in Table~\ref{tab-riemann-tensor}, which are invariant under frame transformation.
We can calculate the frame-independent cutoff scale in the same way as previous sections, although the results are now more involved. 
As an illustration, we compare our results to those without gravity in the limit $v \to 0$,
where the Riemann tensor respects the O$(N)$ symmetry $\bar{R}_{\pi^i \pi^k} = \bar{R}_{h \pi^k}$ as expected.
The frame-independent cutoff scale is
\begin{equation}
	\frac{\Lambda_\text{NLSM}}{\Lambda_\text{G}} \sim \sqrt{\frac{N^2}{\bar R \Lambda_\text{G}^2}} \sim
	\sqrt{\frac{N}{N-1}} \qty( \Mpl^2 \overline{G'_0} + \frac{(1 + 6 \Mpl^2 \overline{f'_0})^2}{6} )^{-1/2}.
\end{equation}
The first term corresponds to the result without gravity and the second term is the correction from the non-minimal coupling to gravity.
The latter vanishes for $\Mpl^2 \overline{f'_0} = - 1/6$, \textit{i.e.,} the conformal coupling, as expected.

As a final remark, we consider a general frame transformation
\begin{equation} \label{eq-F-frame}
	\Phi_\text{F}^2 = \Omega_\text{F}^2 \Phi_\text{J}^2, \qquad \Omega_\text{F}^2 = \frac{f}{f_\text{F}},
\end{equation}
where $f_\text{F}$ being an arbitrary function of $h^2$.
The extended-target space metric for $(\varphi^a_\text{F}) = (\Phi_\text{F}, h, \vec{\pi})$ becomes
\begin{equation}\label{eq-NLSM-Fmetric}
	\qty(\hat{G}^{\text{NL}}_{ab}) = \begin{pmatrix}
		\begin{matrix}
			-f_\text{F} & - f_\text{F}' \Phi_\text{F} h  \\
	- f_\text{F}' \Phi_\text{F} h & \frac{\Phi_\text{F}^2 \qty[ \frac{\mathcal{G}_{hh}}{6 \Mpl^2} + h^2 f \qty( \frac{f'^2}{f^2} - \frac{f_\text{F}'^2}{f_\text{F}^2} ) ]}{\Omega_\text{F}^2}
		\end{matrix}
	& \rvline & \bigzero \\[2em]
	\hline
	\bigzero &  \rvline & \frac{\Phi_\text{F}^2 ( \mathcal{G}_{\pi^i \pi^j} )}{6 \Mpl^2 \Omega_\text{F}^2}\
\end{pmatrix}.
\end{equation}
Interestingly, we can obtain an apparently flat target space for $ \phi^i $ by choosing $f_\text{F} $ such that $ \hat{G}_{hh}^\text{NL} = (6\Mpl^2 \Omega_\text{F}^2)^{-1} \Phi_\text{F}^2 $ for given $ f $ and $ \mathcal{G}_{hh} $, and redefining $ \dd\tilde \phi^i \equiv \sqrt{f_\text{F}/f} \dd\phi^i $. 
In other words, \textit{for a given O$(N)$ NLSM, there always exists a certain frame where the target space of $\phi^i$ is completely flat}.\footnote{
	This particular frame is ``the Jordan frame'' for the metric and Einstein--Cartan HI.
}

This observation emphasizes the significance of the inclusion of conformal mode in the discussion of NLSM with gravity.
As we have shown, all the different target-space geometries of $ \phi^i $ in O$(N)$ NLSM are connected by the frame transformation.
Hence, the target space spanned only by $ \phi^i $ is clearly unphysical once we introduce the coupling to gravity.
To tell the difference, we have to consider the geometry of the extended target space including the conformal mode, which is manifestly invariant under the frame transformation or field redefinition.
As we have seen, we only need one straightforward calculation of the curvature of the extended-target space.

\smallskip\noindent\textit{\textbf{Conclusions.\,---\,}} %
We propose a geometric method for calculation of physical quantities, \textit{e.g.}, the unitarity-violation scale, of theories where scalar fields non-conformally coupled with gravity, from which the results are manifestly Weyl-transformation or field-redefinition independent as they should be.
The cutoff scale of multiple-scalar theories is characterized by the geometry of the target space of scalar fields which is invariant manifestly under redundancy description of field redefinition.
However, coupling to gravity introduces a new redundancy, \textit{i.e.}, Weyl transformation or frame choice, which spoils the advantages of the geometric method.
We show that including the conformal mode of the spacetime metric to extend the notion of target space can help regain the merits of geometric method, because the Weyl transformation now becomes simply a field redefinition (of the conformal mode) in the extended geometry which, by definition, does not change the geometric or physical quantities.
We show several examples of frame-independent unitarity-violation scales, such as HI in metric and Einstein--Cartan formalisms (Table~\ref{tab-riemann-tensor}).
These results are consistent with those calculated in either Jordan or Einstein frame in the literature.
We also discuss general NLSM where one can freely choose a frame in which the original target space is flat or curved but the extended geometry and physical quantities are invariant under field redefinition of both conformal mode and scalar fields.


\medskip\noindent\textit{Acknowledgments\,---\,}%
M.\,H.\, was supported by IBS under the project code, IBS-R018-D1.
K.\,K.\, was supported by JSPS KAKENHI Grant-in-Aid for Challenging Research (Exploratory) Grant No. JP23K17687 and National Natural Science Foundation of China (NSFC) under Grant No. 12347103.
K.\,M.\, was supported by MEXT Leading Initiative for Excellent Young Researchers Grant No.\ JPMXS0320200430,
and by JSPS KAKENHI Grant No.\ 	JP22K14044.


\bibliographystyle{JHEP}
\bibliography{manuscript}

\providecommand{\href}[2]{#2}\begingroup\raggedright\begin{thebibliography}{10}

\bibitem{Meetz:1969as}
K.~Meetz, \emph{{Realization of chiral symmetry in a curved isospin space}}, \href{https://doi.org/10.1063/1.1664881}{\emph{J. Math. Phys.} {\bfseries 10} (1969) 589}.

\bibitem{Honerkamp:1971sh}
J.~Honerkamp, \emph{{Chiral multiloops}}, \href{https://doi.org/10.1016/0550-3213(72)90299-4}{\emph{Nucl. Phys. B} {\bfseries 36} (1972) 130}.

\bibitem{Honerkamp:1971xtx}
J.~Honerkamp and K.~Meetz, \emph{{Chiral-invariant perturbation theory}}, \href{https://doi.org/10.1103/PhysRevD.3.1996}{\emph{Phys. Rev. D} {\bfseries 3} (1971) 1996}.

\bibitem{Ecker:1971xko}
G.~Ecker and J.~Honerkamp, \emph{{Application of invariant renormalization to the nonlinear chiral invariant pion lagrangian in the one-loop approximation}}, \href{https://doi.org/10.1016/0550-3213(71)90468-8}{\emph{Nucl. Phys. B} {\bfseries 35} (1971) 481}.

\bibitem{Alvarez-Gaume:1981exa}
L.~Alvarez-Gaume, D.~Z. Freedman and S.~Mukhi, \emph{{The Background Field Method and the Ultraviolet Structure of the Supersymmetric Nonlinear Sigma Model}}, \href{https://doi.org/10.1016/0003-4916(81)90006-3}{\emph{Annals Phys.} {\bfseries 134} (1981) 85}.

\bibitem{Alvarez-Gaume:1981exv}
L.~Alvarez-Gaume and D.~Z. Freedman, \emph{{Geometrical Structure and Ultraviolet Finiteness in the Supersymmetric Sigma Model}}, \href{https://doi.org/10.1007/BF01208280}{\emph{Commun. Math. Phys.} {\bfseries 80} (1981) 443}.

\bibitem{Boulware:1981ns}
D.~G. Boulware and L.~S. Brown, \emph{{SYMMETRIC SPACE SCALAR FIELD THEORY}}, \href{https://doi.org/10.1016/0003-4916(82)90192-0}{\emph{Annals Phys.} {\bfseries 138} (1982) 392}.

\bibitem{Howe:1986vm}
P.~S. Howe, G.~Papadopoulos and K.~S. Stelle, \emph{{The Background Field Method and the Nonlinear $\sigma$ Model}}, \href{https://doi.org/10.1016/0550-3213(88)90379-3}{\emph{Nucl. Phys. B} {\bfseries 296} (1988) 26}.

\bibitem{Coleman:1969sm}
S.~R. Coleman, J.~Wess and B.~Zumino, \emph{{Structure of phenomenological Lagrangians. 1.}}, \href{https://doi.org/10.1103/PhysRev.177.2239}{\emph{Phys. Rev.} {\bfseries 177} (1969) 2239}.

\bibitem{Callan:1969sn}
C.~G. Callan, Jr., S.~R. Coleman, J.~Wess and B.~Zumino, \emph{{Structure of phenomenological Lagrangians. 2.}}, \href{https://doi.org/10.1103/PhysRev.177.2247}{\emph{Phys. Rev.} {\bfseries 177} (1969) 2247}.

\bibitem{Dixon:1989fj}
L.~J. Dixon, V.~Kaplunovsky and J.~Louis, \emph{{On Effective Field Theories Describing (2,2) Vacua of the Heterotic String}}, \href{https://doi.org/10.1016/0550-3213(90)90057-K}{\emph{Nucl. Phys. B} {\bfseries 329} (1990) 27}.

\bibitem{Alonso:2015fsp}
R.~Alonso, E.~E. Jenkins and A.~V. Manohar, \emph{{A Geometric Formulation of Higgs Effective Field Theory: Measuring the Curvature of Scalar Field Space}}, \href{https://doi.org/10.1016/j.physletb.2016.01.041}{\emph{Phys. Lett. B} {\bfseries 754} (2016) 335} [\href{https://arxiv.org/abs/1511.00724}{{\ttfamily 1511.00724}}].

\bibitem{Alonso:2016oah}
R.~Alonso, E.~E. Jenkins and A.~V. Manohar, \emph{{Geometry of the Scalar Sector}}, \href{https://doi.org/10.1007/JHEP08(2016)101}{\emph{JHEP} {\bfseries 08} (2016) 101} [\href{https://arxiv.org/abs/1605.03602}{{\ttfamily 1605.03602}}].

\bibitem{Nagai:2019tgi}
R.~Nagai, M.~Tanabashi, K.~Tsumura and Y.~Uchida, \emph{{Symmetry and geometry in a generalized Higgs effective field theory: Finiteness of oblique corrections versus perturbative unitarity}}, \href{https://doi.org/10.1103/PhysRevD.100.075020}{\emph{Phys. Rev. D} {\bfseries 100} (2019) 075020} [\href{https://arxiv.org/abs/1904.07618}{{\ttfamily 1904.07618}}].

\bibitem{Cohen:2021ucp}
T.~Cohen, N.~Craig, X.~Lu and D.~Sutherland, \emph{{Unitarity violation and the geometry of Higgs EFTs}}, \href{https://doi.org/10.1007/JHEP12(2021)003}{\emph{JHEP} {\bfseries 12} (2021) 003} [\href{https://arxiv.org/abs/2108.03240}{{\ttfamily 2108.03240}}].

\bibitem{Cheung:2021yog}
C.~Cheung, A.~Helset and J.~Parra-Martinez, \emph{{Geometric soft theorems}}, \href{https://doi.org/10.1007/JHEP04(2022)011}{\emph{JHEP} {\bfseries 04} (2022) 011} [\href{https://arxiv.org/abs/2111.03045}{{\ttfamily 2111.03045}}].

\bibitem{Helset:2022tlf}
A.~Helset, E.~E. Jenkins and A.~V. Manohar, \emph{{Geometry in scattering amplitudes}}, \href{https://doi.org/10.1103/PhysRevD.106.116018}{\emph{Phys. Rev. D} {\bfseries 106} (2022) 116018} [\href{https://arxiv.org/abs/2210.08000}{{\ttfamily 2210.08000}}].

\bibitem{Alonso:2023upf}
R.~Alonso, \emph{{A primer on Higgs Effective Field Theory with Geometry}},  \href{https://arxiv.org/abs/2307.14301}{{\ttfamily 2307.14301}}.

\bibitem{Cervantes-Cota:1995ehs}
J.~L. Cervantes-Cota and H.~Dehnen, \emph{{Induced gravity inflation in the standard model of particle physics}}, \href{https://doi.org/10.1016/0550-3213(95)00128-X}{\emph{Nucl. Phys. B} {\bfseries 442} (1995) 391} [\href{https://arxiv.org/abs/astro-ph/9505069}{{\ttfamily astro-ph/9505069}}].

\bibitem{Bezrukov:2007ep}
F.~L. Bezrukov and M.~Shaposhnikov, \emph{{The Standard Model Higgs boson as the inflaton}}, \href{https://doi.org/10.1016/j.physletb.2007.11.072}{\emph{Phys. Lett. B} {\bfseries 659} (2008) 703} [\href{https://arxiv.org/abs/0710.3755}{{\ttfamily 0710.3755}}].

\bibitem{Barvinsky:2008ia}
A.~O. Barvinsky, A.~Y. Kamenshchik and A.~A. Starobinsky, \emph{{Inflation scenario via the Standard Model Higgs boson and LHC}}, \href{https://doi.org/10.1088/1475-7516/2008/11/021}{\emph{JCAP} {\bfseries 11} (2008) 021} [\href{https://arxiv.org/abs/0809.2104}{{\ttfamily 0809.2104}}].

\bibitem{Planck:2018jri}
{\scshape Planck} collaboration, \emph{{Planck 2018 results. X. Constraints on inflation}}, \href{https://doi.org/10.1051/0004-6361/201833887}{\emph{Astron. Astrophys.} {\bfseries 641} (2020) A10} [\href{https://arxiv.org/abs/1807.06211}{{\ttfamily 1807.06211}}].

\bibitem{Maeda:1988ab}
K.-i. Maeda, \emph{{Towards the Einstein-Hilbert Action via Conformal Transformation}}, \href{https://doi.org/10.1103/PhysRevD.39.3159}{\emph{Phys. Rev. D} {\bfseries 39} (1989) 3159}.

\bibitem{Ema:2016dny}
Y.~Ema, R.~Jinno, K.~Mukaida and K.~Nakayama, \emph{{Violent Preheating in Inflation with Nonminimal Coupling}}, \href{https://doi.org/10.1088/1475-7516/2017/02/045}{\emph{JCAP} {\bfseries 02} (2017) 045} [\href{https://arxiv.org/abs/1609.05209}{{\ttfamily 1609.05209}}].

\bibitem{Hertzberg:2010dc}
M.~P. Hertzberg, \emph{{On Inflation with Non-minimal Coupling}}, \href{https://doi.org/10.1007/JHEP11(2010)023}{\emph{JHEP} {\bfseries 11} (2010) 023} [\href{https://arxiv.org/abs/1002.2995}{{\ttfamily 1002.2995}}].

\bibitem{Bezrukov:2010jz}
F.~Bezrukov, A.~Magnin, M.~Shaposhnikov and S.~Sibiryakov, \emph{{Higgs inflation: consistency and generalisations}}, \href{https://doi.org/10.1007/JHEP01(2011)016}{\emph{JHEP} {\bfseries 01} (2011) 016} [\href{https://arxiv.org/abs/1008.5157}{{\ttfamily 1008.5157}}].

\bibitem{Bezrukov:2011sz}
F.~Bezrukov, D.~Gorbunov and M.~Shaposhnikov, \emph{{Late and early time phenomenology of Higgs-dependent cutoff}}, \href{https://doi.org/10.1088/1475-7516/2011/10/001}{\emph{JCAP} {\bfseries 10} (2011) 001} [\href{https://arxiv.org/abs/1106.5019}{{\ttfamily 1106.5019}}].

\bibitem{Ren:2014sya}
J.~Ren, Z.-Z. Xianyu and H.-J. He, \emph{{Higgs Gravitational Interaction, Weak Boson Scattering, and Higgs Inflation in Jordan and Einstein Frames}}, \href{https://doi.org/10.1088/1475-7516/2014/06/032}{\emph{JCAP} {\bfseries 06} (2014) 032} [\href{https://arxiv.org/abs/1404.4627}{{\ttfamily 1404.4627}}].

\bibitem{Ema:2019fdd}
Y.~Ema, \emph{{Dynamical Emergence of Scalaron in Higgs Inflation}}, \href{https://doi.org/10.1088/1475-7516/2019/09/027}{\emph{JCAP} {\bfseries 09} (2019) 027} [\href{https://arxiv.org/abs/1907.00993}{{\ttfamily 1907.00993}}].

\bibitem{Hill:2020oaj}
C.~T. Hill and G.~G. Ross, \emph{{Gravitational Contact Interactions and the Physical Equivalence of Weyl Transformations in Effective Field Theory}}, \href{https://doi.org/10.1103/PhysRevD.102.125014}{\emph{Phys. Rev. D} {\bfseries 102} (2020) 125014} [\href{https://arxiv.org/abs/2009.14782}{{\ttfamily 2009.14782}}].

\bibitem{Ito:2021ssc}
A.~Ito, W.~Khater and S.~Rasanen, \emph{{Tree-level unitarity in Higgs inflation in the metric and the Palatini formulation}}, \href{https://doi.org/10.1007/JHEP06(2022)164}{\emph{JHEP} {\bfseries 06} (2022) 164} [\href{https://arxiv.org/abs/2111.05621}{{\ttfamily 2111.05621}}].

\bibitem{Karananas:2022byw}
G.~K. Karananas, M.~Shaposhnikov and S.~Zell, \emph{{Field redefinitions, perturbative unitarity and Higgs inflation}}, \href{https://doi.org/10.1007/JHEP06(2022)132}{\emph{JHEP} {\bfseries 06} (2022) 132} [\href{https://arxiv.org/abs/2203.09534}{{\ttfamily 2203.09534}}].

\bibitem{Burgess:2009ea}
C.~P. Burgess, H.~M. Lee and M.~Trott, \emph{{Power-counting and the Validity of the Classical Approximation During Inflation}}, \href{https://doi.org/10.1088/1126-6708/2009/09/103}{\emph{JHEP} {\bfseries 09} (2009) 103} [\href{https://arxiv.org/abs/0902.4465}{{\ttfamily 0902.4465}}].

\bibitem{Barbon:2009ya}
J.~L.~F. Barbon and J.~R. Espinosa, \emph{{On the Naturalness of Higgs Inflation}}, \href{https://doi.org/10.1103/PhysRevD.79.081302}{\emph{Phys. Rev. D} {\bfseries 79} (2009) 081302} [\href{https://arxiv.org/abs/0903.0355}{{\ttfamily 0903.0355}}].

\bibitem{Ema:2020zvg}
Y.~Ema, K.~Mukaida and J.~van~de Vis, \emph{{Higgs inflation as nonlinear sigma model and scalaron as its $\sigma$-meson}}, \href{https://doi.org/10.1007/JHEP11(2020)011}{\emph{JHEP} {\bfseries 11} (2020) 011} [\href{https://arxiv.org/abs/2002.11739}{{\ttfamily 2002.11739}}].

\bibitem{Ema:2020evi}
Y.~Ema, K.~Mukaida and J.~van~de Vis, \emph{{Renormalization group equations of Higgs-R$^{2}$ inflation}}, \href{https://doi.org/10.1007/JHEP02(2021)109}{\emph{JHEP} {\bfseries 02} (2021) 109} [\href{https://arxiv.org/abs/2008.01096}{{\ttfamily 2008.01096}}].

\bibitem{Burgess:2010zq}
C.~P. Burgess, H.~M. Lee and M.~Trott, \emph{{Comment on Higgs Inflation and Naturalness}}, \href{https://doi.org/10.1007/JHEP07(2010)007}{\emph{JHEP} {\bfseries 07} (2010) 007} [\href{https://arxiv.org/abs/1002.2730}{{\ttfamily 1002.2730}}].

\bibitem{Gorbunov:2018llf}
D.~Gorbunov and A.~Tokareva, \emph{{Scalaron the healer: removing the strong-coupling in the Higgs- and Higgs-dilaton inflations}}, \href{https://doi.org/10.1016/j.physletb.2018.11.015}{\emph{Phys. Lett. B} {\bfseries 788} (2019) 37} [\href{https://arxiv.org/abs/1807.02392}{{\ttfamily 1807.02392}}].

\bibitem{Finn:2019aip}
K.~Finn, S.~Karamitsos and A.~Pilaftsis, \emph{{Frame Covariance in Quantum Gravity}}, \href{https://doi.org/10.1103/PhysRevD.102.045014}{\emph{Phys. Rev. D} {\bfseries 102} (2020) 045014} [\href{https://arxiv.org/abs/1910.06661}{{\ttfamily 1910.06661}}].

\bibitem{Mikura:2021clt}
Y.~Mikura and Y.~Tada, \emph{{On UV-completion of Palatini-Higgs inflation}}, \href{https://doi.org/10.1088/1475-7516/2022/05/035}{\emph{JCAP} {\bfseries 05} (2022) 035} [\href{https://arxiv.org/abs/2110.03925}{{\ttfamily 2110.03925}}].

\bibitem{Bauer:2010jg}
F.~Bauer and D.~A. Demir, \emph{{Higgs-Palatini Inflation and Unitarity}}, \href{https://doi.org/10.1016/j.physletb.2011.03.042}{\emph{Phys. Lett. B} {\bfseries 698} (2011) 425} [\href{https://arxiv.org/abs/1012.2900}{{\ttfamily 1012.2900}}].

\bibitem{Rasanen:2018ihz}
S.~Rasanen, \emph{{Higgs inflation in the Palatini formulation with kinetic terms for the metric}}, \href{https://doi.org/10.21105/astro.1811.09514}{\emph{Open J. Astrophys.} {\bfseries 2} (2019) 1} [\href{https://arxiv.org/abs/1811.09514}{{\ttfamily 1811.09514}}].

\bibitem{Nieh:1981ww}
H.~T. Nieh and M.~L. Yan, \emph{{An Identity in Riemann-cartan Geometry}}, \href{https://doi.org/10.1063/1.525379}{\emph{J. Math. Phys.} {\bfseries 23} (1982) 373}.

\bibitem{Shaposhnikov:2020gts}
M.~Shaposhnikov, A.~Shkerin, I.~Timiryasov and S.~Zell, \emph{{Higgs inflation in Einstein-Cartan gravity}}, \href{https://doi.org/10.1088/1475-7516/2021/10/E01}{\emph{JCAP} {\bfseries 02} (2021) 008} [\href{https://arxiv.org/abs/2007.14978}{{\ttfamily 2007.14978}}].

\bibitem{He:2023vlj}
M.~He, K.~Kamada and K.~Mukaida, \emph{{Quantum corrections to Higgs inflation in Einstein-Cartan gravity}}, \href{https://doi.org/10.1007/JHEP01(2024)014}{\emph{JHEP} {\bfseries 01} (2024) 014} [\href{https://arxiv.org/abs/2308.14398}{{\ttfamily 2308.14398}}].

\end{thebibliography}\endgroup



\newpage
\onecolumngrid
\newpage


\renewcommand{\thesection}{S\arabic{section}}
\renewcommand{\theequation}{S\arabic{equation}}
\renewcommand{\thefigure}{S\arabic{figure}}
\renewcommand{\thetable}{S\arabic{table}}
\setcounter{equation}{0}
\setcounter{figure}{0}
\setcounter{table}{0}
\setcounter{page}{1}


\begin{center}
	\textbf{\large Supplemental Material: Geometry and unitarity of scalar fields coupled to gravity}
\end{center}

In this supplemental material, we provide some explicit calculations in a general O$(N)$ NLSM.
The procedure here can be immediately applied to the specific examples of O$(N)$ NLSM such as the metric and Einstein--Cartan HIs.


\section{Target space metric of O$ (N) $ NLSM}
A general NLSM with O$ (N) $ symmetry has a target space metric in the following form
\begin{equation}
    \dd s^2 = \mathcal{G}_{ij} \dd \phi^i \dd \phi^j = \mathcal{G}_1 \dd \phi^i \dd \phi_i +\mathcal{G}_2 \phi_i \phi_j \dd \phi^i \dd \phi^j,
\end{equation}
where $ \mathcal{G}_1 $ and $ \mathcal{G}_2 $ are two positive-definite functions of $ \phi^2 $. In spherical coordinates $ \phi^i= (\phi_r, \vec{\pi}) $, the metric can be written as
\begin{equation}
    \dd s^2 =\left( \mathcal{G}_1 +\mathcal{G}_2 \phi_r^2 \right) \dd\phi_r^2 +\mathcal{G}_1 \phi_r^2 \dd \Omega_{N-1},
\end{equation}
where the angular part can be written as
\begin{equation}
    \dd\Omega_{N-1} =\frac{1}{v^2}
    \left[ \dd \vec{\pi}^2 + \frac{(\vec{\pi}\cdot \dd \vec{\pi})^2}{(v^2 - \vec{\pi}^2)} \right].
\end{equation}
By redefining $ h^2 \equiv \mathcal{G}_1 \phi_r^2 $, one can show that
\begin{equation}
    \dd s^2 = G(h^2) \dd h^2 + \frac{h^2}{v^2}
    \left[ \dd \vec{\pi}^2 + \frac{(\vec{\pi}\cdot \dd \vec{\pi})^2}{(v^2 - \vec{\pi}^2)} \right],
\end{equation}
where we have compactly defined
\begin{equation}
    G(h^2) \equiv \frac{\mathcal{G}_1^2 +\mathcal{G}_2 h^2}{\left( \mathcal{G}_1+ \frac{h^2}{\mathcal{G}_1} \mathcal{G}_1' \right)^2} .
\end{equation}
One can explicitly check the condition of $G(0) = 1$.

\section{Weyl transformation that flattens the target space $ \mathcal{G}_{ij} $}

As can be seen in Eq.~\eqref{eq-NLSM-Fmetric}, a special Weyl transformation or equivalently a special choice of $ f_\text{F} $ can flatten the target space of a O$ (N) $ NLSM, which requires
\begin{equation}
    \left( \ln f_\text{F} \right)'^2 = \left( \ln f \right)'^2 + \frac{\mathcal{G}_{hh}-1}{6\Mpl^2 h^2 f} .
\end{equation}
One may solve this differential equation for a given model, \textit{i.e.} $ f $ and $ \mathcal{G}_{hh} $, to obtain the conformal factor
\begin{equation}
    f_\text{F} (h^2) = \exp \left[ \int^{h^2} \sqrt{\left( \frac{\partial \ln f(x)}{\partial x} \right)^2 + \frac{\mathcal{G}_{hh}(x)-1}{6\Mpl^2 x f(x)} } \dd x \right].
\end{equation}

\section{Frame independence of $\bar G_{ab}$ and $\bar R_{acbd}$}

Consider a general frame transformation of Eq.~\eqref{eq-F-frame}, which induces the transformation of
\begin{equation}
	\Lambda\indices{^{a'}_{a}} \equiv \overline{\frac{\partial \varphi_\text{J}^{a'}}{\partial \varphi_\text{F}^a}} =
	\begin{pmatrix}
		\begin{matrix}
			\bar{\Omega}_\text{F}^{-1} & -2 \frac{\bar{\Phi}_\text{F}}{\bar{\Omega}_\text{F}^2} \overline{\Omega_\text{F}'} v \\
			0 & 1
		\end{matrix}
		&\rvline &  \bigzero \\
		\hline
		\bigzero & \rvline & \mathbb{1}
	\end{pmatrix}.
\end{equation}
In general, the Riemann tensor obtained in the Jordan frame transforms covariantly as
\begin{equation}
	\bar R^\text{F}_{acbd} = \bar R_{a'c'b'd'} \Lambda\indices{^{a'}_{a}} \Lambda\indices{^{b'}_{b}}\Lambda\indices{^{c'}_{c}} \Lambda\indices{^{d'}_{d}}.
\end{equation}
However, the components of Riemann tensor with at least one index as $ \delta \Phi_\text{J} $ vanish, the matrix $ \Lambda\indices{^{a'}_{a}} $ turns out to trivially be a identity matrix, which leaves the Riemann tensor unchanged under general frame transformation.

The target-space metric in a general frame \eqref{eq-NLSM-Fmetric} at $\bar \varphi^a_\text{F}$ is not diagonal.
One may easily canonically normalized the fields by applying the vielbeins
\begin{equation}
	\bar{\hat G}^\text{NL}_{IJ} = \bar e_I^{~a} \bar e_J^{~b}  \bar{\hat{G}}^\text{NL}_{ab}, \qquad
	\qty(\bar e_I^{~a}) \equiv 	\begin{pmatrix}
		\begin{matrix}
			\frac{1}{\bar f_\text{F}^{1/2}} & -\frac{\sqrt{6} v \overline{f'_\text{F}}\bar f^{1/2}}{\bar{f}_\text{F}^{3/2}} \frac{1}{\qty( \frac{\bar{\mathcal{G}}_{hh}}{\Mpl^2} + 6v^2 \frac{\overline{f'}^2}{\bar f})^{1/2}} \\
			0 & \frac{\bar f^{1/2}}{\Lambda_\text{G}} \frac{1}{\qty( \frac{\bar{\mathcal{G}}_{hh}}{\Mpl^2} + 6v^2 \frac{\overline{f'}^2}{\bar f})^{1/2}}
		\end{matrix}
		&\rvline &  \bigzero \\
		\hline
		\bigzero & \rvline & \frac{\Mpl}{\Lambda_\text{G}} \bar{f}^{1/2} \mathbb{1}
	\end{pmatrix}.
\end{equation}

\end{document}